\documentclass{aa}
\usepackage{graphics,epsfig}

\newcommand{\kms}{km~s$^{-1}$}
\newcommand{\dV}{\mathrm{d} V}
\newcommand{\cm}{cm$^{-2}$}

\newcommand{\dla}{ damped Lyman-$\alpha$ }
\newcommand{\NHI}{N_{\rm HI}}
\newcommand{\noi}{\noindent}

\newcommand{\acmsqr}{atoms~cm$^{-2}$}
\newcommand{\lya}{Lyman-$\alpha\ $}

\begin{document}
\title{A deep search for 21cm absorption in high redshift damped Lyman-$\alpha$ systems.}
\titlerunning{21cm absorption in high $z$ DLAs}
\author{Nissim Kanekar \inst{1}\thanks{nissim@ncra.tifr.res.in},
Jayaram N Chengalur\inst{2}\thanks{chengalu@ncra.tifr.res.in}}
\authorrunning{Kanekar \& Chengalur }
\institute{Kapteyn Institute, University of Groningen, Post Bag 800, 9700 AV Groningen
\and National Centre for Radio Astrophysics, Post Bag 3, Ganeshkhind, Pune 411 007 }
\date{Received mmddyy/ accepted mmddyy}
\offprints{Nissim Kanekar}
\abstract{We present deep GMRT 21cm absorption spectra of 10 damped Lyman-$\alpha$ 
systems (DLAs), of which 8 are at redshifts $z \ga 1.3$. HI absorption was detected 
in only one DLA, the $z = 0.5318$ absorber toward PKS~1629+12. This absorber has been
identified with a luminous spiral galaxy; the spin temperature limit ($T_{\rm s} \le 310$~K)
derived from our observations continues the trend of DLAs associated with bright spirals
having low spin temperatures. In seven of the remaining 9 systems, the observations place 
strong lower limits on the spin temperature of the HI gas. \\
	We combine this sample with data taken from the literature to study the
properties of all known DLAs with 21cm absorption studies. The sample of DLAs which have 
been searched for 21cm absorption now consists of 31 systems, with $T_{\rm s}$ estimates 
available in 24 cases; of these, 16 are at $z < 2$ and 8 at $z > 2$, with 11 (all at $z < 1$) 
having optical identifications. For the latter 11 DLAs, we find that all of the 
low $T_{\rm s}$ DLAs have been identified with large, luminous galaxies, while all the DLAs 
with high spin temperature ($T_{\rm s} \ga 1000$~K) have been identified either with LSBs 
or dwarfs. Further, we find no correlation between impact parameter and spin 
temperature; it is thus unlikely  that the high measured $T_{\rm s}$ values for DLAs 
arise from lines of sight passing through the outskirts of large disk galaxies. 
Instead, the spin temperature of DLAs appears to correlate with the host galaxy type.\\
	The trend (noted earlier by Chengalur \& Kanekar 2000) that low $z$ DLAs 
exhibit both high and low $T_{\rm s}$ values while high redshift ($z \ga 3$) DLAs only show high 
spin temperatures is present in this expanded data set. Based on this difference in
spin temperatures, the Gehan test rules out the hypothesis that DLAs at $z > 2$ and 
DLAs at $z < 2$ are drawn  from the same parent population at $\sim 99$\% confidence 
level. \\
	Finally, we use the new GMRT spectra along with 2 spectra from the literature
to estimate upper limits on the fraction of cold HI, $f_{\rm CNM}$, in DLAs at $z \ga 3$.
For local spirals, $f_{\rm CNM} \sim 0.5$; in contrast, we find that $f_{\rm CNM} < 0.3$ 
in all 7 high $z$ absorbers, and $f_{\rm CNM} < 0.1$ in 5 of the 7 cases. 
}
\maketitle

\section{Introduction}
	
	While molecular gas can be detected in emission at high redshifts 
(albeit only in a handful of somewhat unusual systems), it is 
as yet a severe challenge to detect neutral atomic gas in emission, 
even at redshifts as low as $z \sim 0.2$. As a result, absorption spectra 
against bright background sources provide the sole source of information on 
neutral atomic gas in the high redshift Universe. An important advantage of using 
absorption spectra as a probe of systems lying along the lines of sight to 
distant QSOs is that such studies tend to be unbiased, at least to first order --
sources which give rise to absorption are presumably typical systems at 
that redshift. In contrast, emission studies of a flux-limited, high redshift
sample are usually dominated by the brightest (often atypical) sources.

	The large absorption cross-section of the \lya transition and the
dominance of hydrogen in baryonic material makes the \lya line one of the 
easiest to detect in QSO spectra. In fact, lines arising from neutral gas with 
column densities as low as $10^{13}$~\acmsqr~can be routinely detected with
present day telescopes. However, although these lines (the \lya ``forest'') 
are by far the most numerous, they contain only a small fraction of the total 
neutral gas at high redshifts; most of the neutral gas is, in fact, contained 
in relatively rare, high column density ($\NHI \ga 10^{20}$~\acmsqr) systems, 
the so-called \dla absorbers (DLAs). As the major repository of neutral gas at 
high redshift, these systems are natural candidates for the precursors of today's
galaxies. However, despite their importance in the context of galaxy evolution, 
the nature of high redshift DLAs is presently quite controversial, with models 
ranging from large rapidly rotating disks (\cite{wolfe86}; \cite{pw97}) to merging 
sub-galactic blobs (\cite{haehnelt98}) to outflows from dwarf galaxies 
(\cite{schaye01}) present in the recent literature.\\
	A successful model for DLAs needs to not only reproduce the 
observed $\Omega_{\rm g}(z)$ and column density distribution but also to 
describe the observed physical conditions in the absorbing gas. The latter
is important for our understanding of DLAs, quite independent of their precise 
size and structure. Conversely, knowledge of these conditions is useful in 
modelling the further temporal evolution of damped systems. HI 21cm observations (possible 
only for those DLAs for which the background quasar is also radio loud) 
are particularly useful because they allow one to estimate the average 
spin temperature $T_{\rm s}$ of the absorbing gas. DLA spin temperatures have been, 
in general, found to be significantly higher than those typically seen 
in lines of sight through the Galactic disk, or in those through the
disks of nearby galaxies (see e.g. \cite{chengalur2000}, \cite{kanekar2001c} 
and references therein). The only exceptions to the above are DLAs which 
have been identified as associated with low redshift ($z~<~0.6$) spiral galaxies 
-- these systems all have low spin temperatures, comparable to those seen 
in the Milky Way and nearby spirals. Unfortunately, the number of systems 
for  which deep 21cm observations are available is relatively small; further, 
while the present data (\cite{chengalur2000}) suggest that high $z$ DLAs 
are different from low $z$ ones, the observations are currently biased toward 
systems at $z< 1$, with not too many systems of the sample at high redshift. 
Are high redshift DLAs indeed systematically different from low redshift ones ? 
Also, how significant is the fact that only DLAs associated with spiral galaxies 
have low spin temperatures ? To find answers to these and similar questions,
it is crucial to increase the number of systems with 21cm observations, 
particularly at high redshifts. We present, in this paper, deep Giant 
Meterwave Radio Telescope (GMRT) observations of 10 DLAs, 8 of which are 
at $z>1$. Two of these systems, the $z = 2.9084$ DLA towards TXS~2342+342 
and the $z = 3.0619$ absorber towards PKS~0336$-$014 have been earlier observed 
by Carilli et al. (1996); the present observations are far more sensitive
than the earlier ones. The observations themselves are summarised in 
Sect.~\ref{sec:obs}, while the results are presented in Sect.~\ref{sec:res} and their 
implications for physical conditions in DLAs discussed in Sect.~\ref{sec:dis}.

\section{Observations and Data Analysis}
\label{sec:obs}
\subsection{The GMRT observations}
\label{ssec:gmrt}

The GMRT has five currently operational frequency bands, at 150~MHz, 233~MHz, 327~MHz, 
610~MHz and 1420~MHz.  The present observations utilised the 327, 610 and 1420~MHz bands, 
which cover redshift ranges $z  \sim 2.9 - 3.6$, $ z \sim 1.15  - 1.5$ and $ z \sim 
0 - 0.6$ respectively, with reasonable sensitivity (i.e. better than half the sensitivity 
at the band centre).  The observations were carried out between January and November 2001, 
using the 30-station FX correlator as the backend; this gives a fixed number of 128 
channels over a bandwidth which can be varied between 64~kHz and 16~MHz. All our 
observations (except for those of PKS~2128$-$123 (bandwidth = 2~MHz) and PKS~1629+12, 
which are described below) used a bandwidth of 1~MHz, sub-divided into 128 channels; 
this gave a channel resolution of 7.8~kHz. The number of available antennas varied 
between 18 and 26, due to various debugging and maintenance activities. One (or more) 
of the standard calibrators 3C48, 3C147 or 3C286 was used in all cases to calibrate 
the absolute flux scale and the system bandpass; these sources were observed at least 
every two hours in all cases, to ensure a good bandpass calibration.  A nearby 
standard phase calibrator (see Cols.~4 and 5 of Table~1) was observed every forty 
minutes in all cases to determine the instrumental phase, except in the case of 
PKS~2128$-$123 which is itself a phase calibrator for the GMRT. Our experience with 
the GMRT indicates that the flux calibration is reliable to $\sim 15$\%, in this
observing mode. Observational details 
are summarised in Table~1, where the sources are arranged in order of increasing 
redshift; note that the values listed here for spectral resolution and RMS 
noise are before any smoothing. 

The only source towards which we detected 21cm absorption, viz. PKS~1629+12, was initially 
observed with a 1 MHz bandwidth on the 12th and 22nd of October 2001, 
with 20 antennas and a total on-source time of $\sim 6$~hours. These observations resulted in 
the detection of a narrow 21cm absorption feature; the source was hence re-observed on the 17th
and 18th of November, 2001 with a 0.25~MHz bandwidth (yielding a resolution of 1.95~kHz,
i.e. 0.6~\kms), to try to resolve out the absorption line. The total on-source time
of these latter observations was 13~hours, with 25 antennas. In all cases, 3C286 was
used for absolute flux and system bandpass calibration while the compact source
1640+123 was used for phase calibration.

\begin{table*}
\begin{center}
\label{tab:obs}
\caption{Observing details}
\begin{tabular}{|c|c|c|c|c|c|c|c|c|c|} \hline
&&&&&&&&&\\
 QSO &        $z_{\rm abs}$ & $\nu_{\rm obs}$ & Phase & Phase Cal. & No. of & Time & Source & Channel & RMS    \\
     &                  & MHz   & Cal.  & Flux &antennas& Hrs. & Flux & resolution& mJy         \\
     &                  &       &       & Jy   & 	&      & Jy   & \kms & \\
&&&&&&&&&\\
\hline
&&&&&&&&&\\
PKS~2128$-$123 & 0.4298 & 993.43 &$^a$       &$^a$ & 20 & 8   & 1.90    & 4.7   & 1.4      \\
&&&&&&&&&\\
PKS~1629+12    & 0.5318 & 927.28 &1640+123  & 2.5 & 25 & 13  & 2.35    & 0.6   & 2.8      \\
&&&&&&&&&\\
PKS~0215+015   & 1.3439 & 605.72 & 0204+152 & 5.0 & 26 & 5   & 0.92    & 3.9   & 2.2     \\
&&&&&&&&&\\
QSO~0957+561A  & 1.3911 & 594.04 & 0834+555 & 7.4 & 26 & 6.5 & 0.59    & 3.9   & 2.4     \\
&&&&&&&&&\\
PKS~1354+258   & 1.4205 & 586.82 & 3C286    & $^b$& 22 & 5.5 & 0.30$^c$& 4.0   & 2.8      \\
&&&&&&&&&\\
TXS~2342+342   & 2.9084 & 363.42 & 3C19     & 8.3 & 23 & 8   & 0.31    & 6.4   & 2.8      \\
&&&&&&&&&\\
PKS~1354$-$107 & 2.966  & 358.15 & 3C283    & 21  & 22 & 9.5 & 0.12    & 6.5   & 2.0      \\
&&&&&&&&&\\
PKS~0537$-$286 & 2.974  & 357.43 &0521$-$207& 7.2 & 25 & 5   & 1.05    & 6.5   & 2.4     \\
&&&&&&&&&\\
PKS~0336$-$014 & 3.0619 & 349.69 & 0323+055 & 7.4 & 26 &  5  & 0.94    & 6.7   & 2.1     \\
&&&&&&&&&\\
PKS~0335$-$122 & 3.178  & 339.97 &0409$-$179& 6.1 & 18 & 5.5 & 0.68    & 6.9   & 1.9     \\
&&&&&&&&&\\
\hline
\end{tabular}
\end{center}
\vskip 0.1 in
${}$~$^a$~No phase calibrator was used, as PKS~2128$-$123 is itself a phase calibrator 
for the GMRT. \\
${}$~$^b$~3C286 was also used as the flux calibrator; we hence do not have an independent 
estimate of its flux.\\
${}$~$^c$~PKS~1354+258 is resolved by the GMRT synthesised beam; this is the peak flux.\\
\vskip 0.1 in

\end{table*}

\subsection{Data Analysis}
\label{ssec:anal}

The data were analysed in classic AIPS using standard procedures. After the initial flagging 
and gain and bandpass calibration, continuum images were made for all sources and 
then used to self-calibrate the U-V data. This was carried out in an iterative manner 
until the quality of the image was found to not improve on further self-calibration. 
In general, a few rounds of phase self-calibration were carried out on all fields,
followed by one or two rounds of amplitude self-calibration. The continuum emission 
was then subtracted from the multi-channel U-V data set, using the AIPS task UVSUB; 
any residual continuum was subtracted by fitting a linear baseline to the U-V 
visibilities, using the task UVLIN. The continuum subtracted data were then mapped 
in all channels and a spectrum extracted at the quasar location from the resulting
three dimensional data cube. In the case of PKS~1629+12, which showed 21cm absorption, 
spectra were also extracted from other locations in the cube to ensure that the 
data were not corrupted by RFI. In general, all the observing frequencies were 
found to be reasonably free of RFI, except for sporadic interference on individual 
antennas which was edited out. In the case of multi-epoch observations 
of a single source, the spectra were shifted to the heliocentric frame outside AIPS 
and then averaged together. 

\section{Results}
\label{sec:res}

\begin{figure*}[t!]
\begin{center}
\epsfig{file=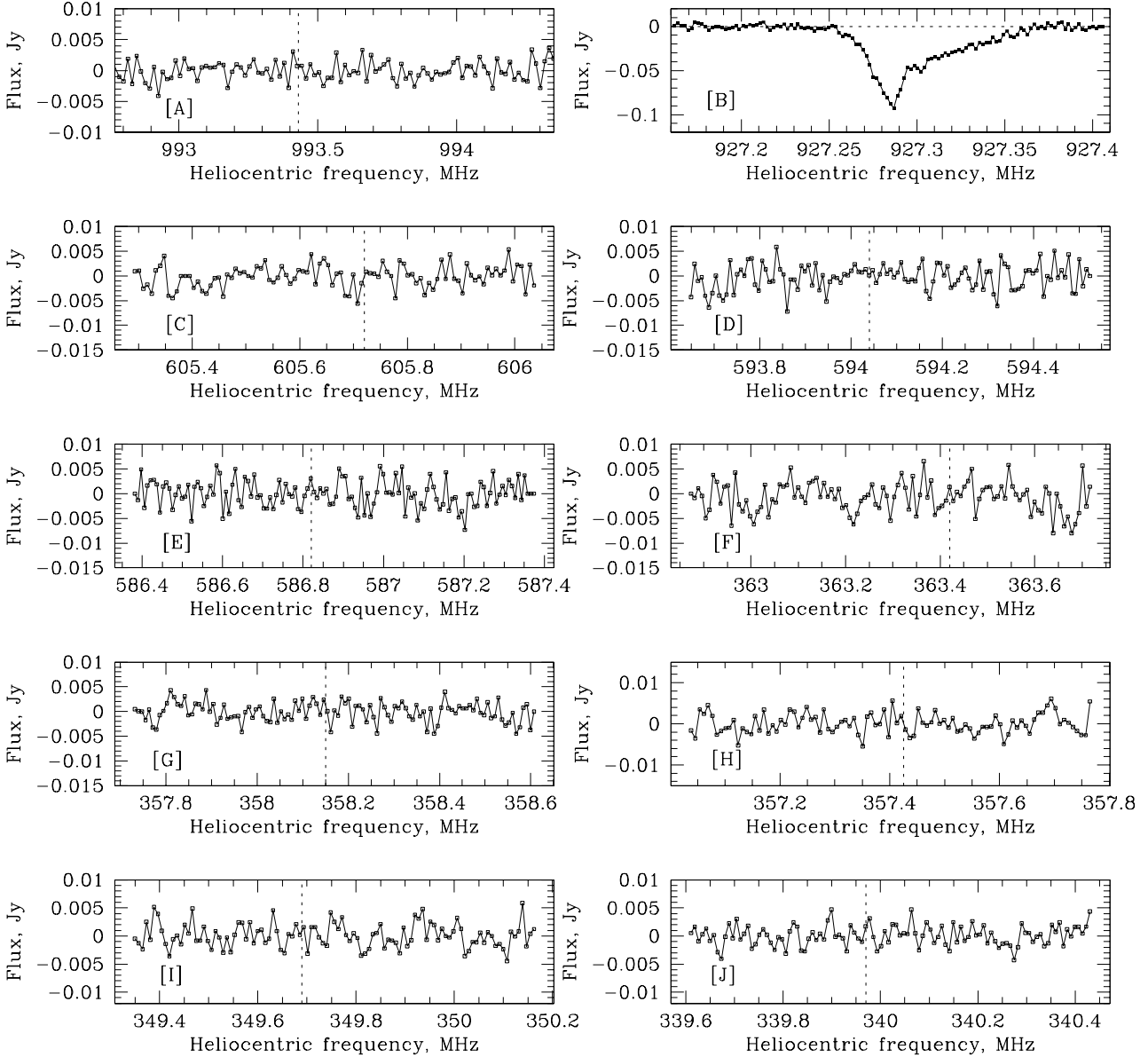,height=7truein}
\end{center}
\caption{ GMRT HI absorption spectra towards the ten sources of our sample, 
arranged in order of increasing redshift. 
{\bf [A]}~$z = 0.4298$~DLA towards PKS~2128$-$123.
{\bf [B]}~$z = 0.5318$~DLA towards PKS~1629+12
{\bf [C]}~$z = 1.3449$~DLA towards PKS~0215+015. 
{\bf [D]}~$z = 1.3911$~DLA towards QSO~0957+561A. 
{\bf [E]}~$z = 1.4205$~ DLA towards PKS~1354+258. 
{\bf [F]}~$z = 2.9084$~DLA towards TXS~2342+342. 
{\bf [G]}~$z = 2.966$~DLA towards PKS~1354$-$107. 
{\bf [H]}~$z = 2.974$~DLA towards PKS~0537$-$286
{\bf [I]}~$z = 3.0619$~DLA towards PKS~0336$-$014. 
{\bf [J]}~$z = 3.178$~DLA towards PKS~0335$-$122.}
\label{fig:spc}
\end{figure*}

The final absorption spectra towards the ten sources of our sample are shown 
in Fig.~\ref{fig:spc}, arranged in order of increasing redshift; the measured 
source fluxes, channel resolutions (in \kms) and RMS noise values are listed in 
Cols.~8, 9 and 10 of Table~1. 

\begin{figure}
\centering
\epsfig{file=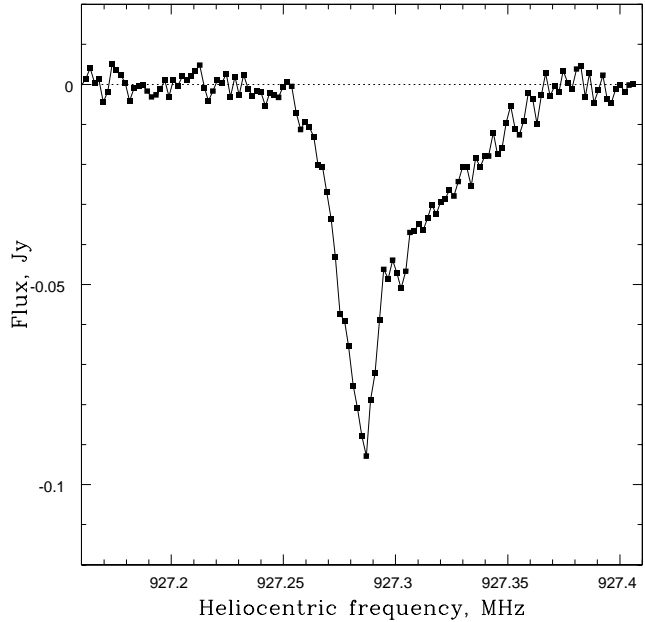,height=3.5truein,width=3.5truein}
\caption{ High resolution (0.6~\kms) GMRT HI absorption spectrum towards 
PKS~1629+12. Asymmetric HI absorption is clearly visible, with the peak 
optical depth at $z = 0.531787$.}
\label{fig:1629}
\end{figure}

Our only detection of HI absorption is in the $z = 0.5318$ DLA towards PKS~1629+12. 
Fig.~\ref{fig:1629} shows a more detailed version of the GMRT 0.25 MHz HI spectrum towards 
this source; the spectrum has a resolution of $\sim 0.6$~\kms~and an RMS noise of 
2.8~mJy. Highly asymmetric HI absorption can be clearly seen, extending 
over a velocity width of $\sim 40$~\kms. The peak absorption occurs at a heliocentric 
frequency of 927.287~MHz, i.e. at a heliocentric redshift of $ 0.531787 \pm 0.000002$. 
This lies in between (and in reasonable agreement with) the redshifts of the 
damped Lyman-$\alpha$ line ($z_{\rm DLA} = 0.532$; \cite{nestor2001}) and the MgI 
absorption ($z_{\rm MgI} = 0.5316$; (\cite{aldcroft94}; \cite{barthel90}).
(Note that Aldcroft et al. (1994) quote $z = 0.5313$ for the strong MgII and FeII 
absorption lines.) The flux of PKS~1629+12 was measured to be 2.35~Jy; the peak optical 
depth is thus $\tau_{\rm max} = 0.039$. The equivalent width of the profile is 
$\int \tau \dV = 0.494 \pm 0.001$~\kms, obtained by integrating the observed absorption profile.

\section{The spin temperature}
\label{sec:Ts}

\begin{table*}
\begin{center}
\label{tab:ts_obs}
\caption{The spin temperature for the 10 DLAs observed here}
\begin{tabular}{|c|c|c|c|c|c|c|c|c|} \hline
&&&&&&&&\\
 QSO &        $z_{\rm abs}$ & $\NHI \times 10^{20}$ & $\Delta V$~$^a$  & RMS$^c$ & $\tau_{\rm max}^d$ 
 &  $(1/f)\int \tau {\mathrm d}V^e$ & $ T_{\rm s}$ & Refs.$^f$ \\
     &                  & \cm  &   \kms     & mJy &        & \kms    & K     &  \\
&&&&&&&&\\
\hline
&&&&&&&&\\
PKS~2128$-$123   & 0.4298 & $0.25\pm0.06$ & 7    & 1.2 & $< 0.0019$ & $< 0.014$ & $ > 980 $ &  1 \\ 
&&&&&&&&\\
PKS~1629+12      & 0.5318 & 2.8  & 0.6  & 2.8 &  0.039     & $0.494 \pm 0.001$   & 310       &  2 \\
&&&&&&&&\\
PKS~0215+015     & 1.3439 & 0.8  & 7.8  & 1.6 & $< 0.0052$ & $< 0.043$ & $ > 1020$ &  3 \\
&&&&&&&&\\
QSO~0957+561A$^b$& 1.3911 & $2.1\pm0.5$ & --   &  -- &    --      & --        & --        &  4 \\
&&&&&&&&\\
PKS~1354+258$^b$ & 1.4205 & $32\pm2$   &  --  &  -- &    --      & --        & --        &  4 \\
&&&&&&&&\\
TXS~2342+342     & 2.9084 & $20\pm0.5$   & 15   & 2.0 & $< 0.0192$ & $< 0.306$ & $ > 3585$ &  5 \\ 
&&&&&&&&\\
PKS~1354$-$107   & 2.966  & 6.0  & 6.5  & 2.0 & $< 0.05$   & $< 0.345$ & $ > 955 $ &  6 \\
&&&&&&&&\\
PKS~0537$-$286   & 2.974  & 2.0  & 10   & 1.9 & $< 0.0054$ & $< 0.058$ & $ > 1890 $ &  6 \\
&&&&&&&&\\
PKS~0336$-$014   & 3.0619 & $16\pm1.3$ & 20   & 1.4 & $< 0.0045$ & $< 0.095$ & $ > 9240$ &  7 \\
&&&&&&&&\\
PKS~0335$-$122   & 3.178  & $6\pm1$  & 15   & 1.2 & $< 0.0053$ & $< 0.084$ & $ > 3920 $ &  6 \\
&&&&&&&&\\
\hline
\end{tabular}
\end{center}
\vskip 0.1 in
${}^a$$\Delta V$ is the FWHM used to compute the spin temperature in the
case of non-detections (see the text for more details). For 1629+12, it is the velocity
resolution of the observations.\\
${}^b$The large uncertainty in the covering factor makes it difficult to
estimate the spin temperature (see the text for more details).\\
${}^c$This is the RMS acually measured from the smoothed spectra. The spectra
were smoothed to a velocity resolution given in the preceeding column, except for
the following cases for which the RMS was measured at a slightly better spectral 
resolution than $\Delta V$, viz 12.8~\kms~for TXS~2342+342, 13.4~\kms~for
PKS~0336$-$014 and  13.8~\kms~for PKS~0335$-$122.\\
${}^d$All limits (in this and other columns) correspond to 3$\sigma$. \\
${}^e$The covering factor $f$ has been set to unity in all cases (see the text for
details).\\
${}^f$References for the HI column densities : 1.~\cite{ledoux2002}; 
2.~\cite{nestor2001}; 3.~\cite{lanzetta95}; 4.~\cite{rao2000}; 5.~\cite{white93}; 
6.~\cite{ellison2001}; 7.~\cite{prochaska01}.
\vskip 0.1 in
\end{table*}

\noi In the case of an optically thin, homogenous cloud in thermal equilibrium, its 
HI column density, $\NHI$, 21cm optical depth, $\tau$, and spin temperature, $T_{\rm s}$, 
are related by the expression (e.g. \cite{rohlfs86})
\begin{equation}
\label{eqn:tspin}
N_{\rm HI} = { 1.823\times10^{18} T_{\rm s} \over f} \int \tau \dV \; ,
\end{equation}
\noi where $N_{\rm HI} $ is in cm$^{-2}$, $T_{\rm s}$ in K and $\dV$ in km s$^{-1}$. $f$ is 
the covering factor of the absorber. Of course, a given line of sight through an 
absorber is quite likely to pass through a number of clouds, with different
spin temperatures and column densities. In such a situation, the spin temperature 
obtained using the above expression is the column density weighted harmonic mean 
of the spin temperatures of the individual clouds (provided, of course, that 
all the clouds are optically thin).

	It should be noted that Eq.~(\ref{eqn:tspin}) relates the HI column density 
to the integrated optical depth $\int \tau {\mathrm d}V$; this means that, 
in the case of a non-detection, one needs to know the shape of the absorption 
profile to compute a lower limit on the spin temperature. This shape is, of course, 
not a priori known. However, since our limits on the optical depth imply that 
the HI line is optically thin, the line can be assumed to have a Gaussian shape; 
here, $\int \tau dV = 1.06 \tau_{\rm max} \Delta V$, where $\tau_{\rm max}$ is the peak 
optical depth (or the limit on the peak) and $\Delta V$ is the FWHM of the Gaussian. 
Further, for a gas cloud in thermal equilibrium, the kinetic temperature $T_{\rm k}$ (with
$T_{\rm s} \approx T_{\rm k}$, for a single cloud) and the FWHM of the observed line are 
related ($T_{\rm k} \approx 21.855 {\Delta V}^2$, with $T_{\rm k}$ in K and $\Delta V$ in 
km~s$^{-1}$); care must hence be taken to derive the limits on $T_{\rm s}$ in a self-consistent 
manner, so that the velocity width over which the integral is carried out is 
consistent with (at least) thermal broadening of the gas at the derived temperature
(any bulk kinematic motion will further broaden the absorption features; we will 
assume that no such motions are present). For example, if the $3 \sigma$ lower limit on 
the spin temperature is 10000~K, one should use $\Delta V \ga 20$~\kms, while, if the 
limit is 1000~K, it is appropriate to use $\Delta V \ga 7$~\kms.  We have, in the 
following analysis, used $\Delta V \approx 7$, $15$ and $20$~\kms~for temperatures 
$T_{\rm s} \sim 1000$, $5000$ and $10000$~K, respectively, and have smoothed the spectra 
to the above resolutions while computing the spin temperature.

The HI column density of a DLA can be directly estimated from the equivalent 
width of the Lyman-$\alpha$ profile. The original criterion for a system to be classified
as a damped absorber ($N_{\rm HI} \ge 2 \times 10^{20}$~\cm) was an observational one, 
linked to the ability to identify damped profiles in medium resolution spectra 
from the relatively small telescopes used to carry out the early DLA surveys. 
Throughout this paper, however, we will consider all systems for which the 
Lyman-$\alpha$ profile shows damping wings as DLAs, regardless of whether or not 
their HI column density is greater than the traditionally used threshold. We hence 
retain the absorbers towards PKS~0215+015 and PKS~2128$-$123 in our DLA sample, 
despite their HI column densities being less than $10^{20}$~\cm.

Given the HI column density, a search for 21cm absorption directly yields the spin 
temperature of the absorbing gas (or limits on the temperature, in the case of 
non-detections of absorption), provided the covering factor $f$ is 
known. Unfortunately, the radio continuum from quasars tends to arise from a 
far more extended region than the optical or UV emission and it may thus transpire 
that some fraction of the radio continuum may ``leak out'' from around the cloud 
and result in an incorrectly estimated spin temperature. Similarly, the line of 
sight along which the 21cm optical depth has been measured need not be the same 
as that for which one has an estimate for the HI column density from the Lyman-$\alpha$ 
line. The fraction of radio emission originating in compact components spatially
coincident with the UV point source can usually, however, be measured from VLBI 
observations (when such observations are available at sufficiently low frequencies).
One can then estimate $f$ and hence, the spin temperature; this is done next, for the 
ten sources of this paper, in order of increasing right ascension. The results are 
summarised in Table~2 in which Cols.~4 to 8 contain, respectively, the velocity 
width $\Delta V$ used to compute $T_{\rm s}$, the spectral RMS over this (or a slightly 
smaller) velocity width, the peak optical depth $\tau_{\rm max}$, 
$(1/f)\int \tau {\mathrm d}V$, and the spin temperature $T_{\rm s}$ (or limits on the 
last three quantities).

1. VLBA maps of PKS~0215+015 at 2.3 and 8.4 GHz (from the Radio Reference Frame Image 
Database (RRFID)) show that the source is exceedingly compact, with most of the 
flux contained within the central $30$~milli-arcseconds. The quasar also has a flat 
spectrum 
between 8.4 GHz and 610 MHz, again implying that it is quite likely to be compact 
even at the lower frequencies. It is hence 
likely that the covering factor $f$ is close to unity. The HI column density of 
the $z = 1.345$ DLA is $\NHI = 8 \times 10^{19}$~\cm~(\cite{lanzetta95}); the 
spin temperature of the absorber is then $T_{\rm s} (3 \sigma) > 1020$~K (using $\Delta V 
= 7.8$~\kms).

2. The $z = 3.178$ DLA towards PKS~0335$-$122 is the highest redshift system of 
our sample, with $\NHI = 6 \pm 1 \times 10^{20}$~\cm~(\cite{ellison2001}). Unfortunately, 
no VLBI information exists in the literature on the background quasar; however, this 
source has a fairly flat spectrum between 5 GHz and 1.4 GHz, with fluxes of 380 mJy 
and 420 mJy at these two frequencies. This spectral flatness implies that the flux 
is likely to arise from a fairly compact source component, which is hence likely 
to be covered by the DLA. This gives a lower limit to the covering factor, 
$f \ge 0.7$, which yields $T_{\rm s} (3 \sigma) > 2740$~K; if $f = 1$, one has $T_{\rm s} 
 (3 \sigma) > 3920$~K ($\Delta V = 15$~\kms).

3. In the case of PKS~0336$-$014, 5 GHz VLBI observations (\cite{gurvits94}) have 
shown that the source is highly compact at this frequency, with the entire 
flux contained within the central $\sim 7$~milli-arcseconds. However, the source 
also has a negative spectral index, indicating the presence of some extended emission. 
In the absence of 
VLBI information at lower frequencies, we will assume a covering factor of unity.
The HI column density of the $z = 3.0619$ DLA is $\NHI = 1.6 \pm 0.13 \times 10^{21}$~\cm~
(\cite{prochaska01}) giving a spin temperature $T_{\rm s} (3 \sigma) > 9240$~K ($\Delta V = 
20$~\kms). Carilli et al.  (1996) estimated $T_{\rm s} > 2500$~K for this DLA.

4. PKS~0537$-$286 has an exceedingly flat spectrum between 8.4~GHz and 327~MHz; for 
example, the QSO flux density was measured to be 0.9~Jy at 8~GHz (Parkes survey, 
\cite{wright90}), 1.2~Jy at 4.85~GHz (\cite{griffith94}), 838~mJy at 
1.4~GHz (NRAO-VLA Sky Survey, \cite{condon98}) and 1.05~Jy in our GMRT 
327~MHz image. This implies that the source is likely to be highly compact. 
VLBA observations at 5~GHz (\cite{fomalont00}) have shown that the flux within 
the central 10~milli-arcseconds is $\sim 500$~mJy, implying a covering factor 
greater than 0.5. In fact, these observations recovered the entire 5~GHz flux of 
1.2~Jy on the shortest VLBA baseline ($2 \times 10^6$ wavelengths, 
\cite{hirabayashi00}), implying that the entire source is contained within 
$\sim  0.1$~arcseconds. It is thus quite likely that the covering factor is close 
to unity; we will use $f = 1$ in this analysis. The HI column density of the 
$z = 2.9742$ DLA has been found to be $2.0 \times 10^{20}$~\cm~(\cite{ellison2001});
we then obtain a $3\sigma$ lower limit of $1890 $~K on the spin temperature of the 
absorbing gas ($\Delta V = 10$~\kms).

5. VLBI mapping of QSO~0957+561 has been carried out on a number of occasions, at
various frequencies at and above 1.6~GHz (see, for example, \cite{garrett94,campbell95}).
The flux of component A has been measured to be only about 32~mJy at 1.6~GHz 
(\cite{garrett94}); clearly, most of the flux seen in the 610~MHz GMRT map comes 
from extended emission which is unlikely to be covered by the DLA. The covering 
factor is thus quite uncertain for this absorber, although it is likely to be 
low. Given this, limits on the spin temperature derived from the present data are 
unlikely to be reliable. In fact, it is known that the DLA does not 
cover the second image (B) of this gravitational lens system (\cite{rao2000}).
We will hence not include this system in the later analysis and also do not 
list limits on the spin temperature and the optical depth in Table~2.

6. PKS~1354+258 is also an extended source at 610~MHz, with no VLBI information 
available in the literature; the DLA covering factor is hence again quite uncertain. 
However, the extremely high column density of the absorber ($N_{\rm HI} = 3.2 \pm 0.2
\times 10^{21}$~ cm$^{-2}$; \cite{rao2000}) results in high lower limits on the 
spin temperature even in the case of low covering factors. For example, 
$f = 0.1$ gives $T_{\rm s} (3 \sigma) > 1100$~K, while $f= 1$ would imply $T_{\rm s} (3 \sigma) 
> 7000 $~K (assuming velocity widths of $8$~\kms~and $20$~\kms~respectively. 
We will however again not quote a spin temperature in Table~2 due 
to the uncertainties in the covering factor.

7. No information exists in the literature regarding the VLBI structure of 
PKS~1354$-$107. However, the source has a strongly inverted spectrum and 
is thus likely to be very compact. We will hence assume a covering factor 
of unity. Combining this with the HI column density of $6 \times 10^{20}$~\cm~
(\cite{ellison2001}) gives $T_{\rm s} (3 \sigma) > 955 $~K ($\Delta V = 6.5$~\kms).

8. The HI column density of the $z = 0.5318$ DLA towards PKS~1629+12 is $\NHI = 
2.8 \times 10^{20}$~\cm~(\cite{nestor2001}). PKS~1629+12 is a well-known compact 
steep spectrum source with an extremely one-sided structure (\cite{saikia90}). 
The 2.3 GHz VLBI map of Dallacasa et al. (1998) shows two components with 
fluxes of 110 mJy and 150 mJy. Only the 110~mJy western component is seen in their 
8.4 GHz map, with almost the same flux; they hence identify this as the flat 
spectrum core of the source. The separation between the two components is 
1.14~arcsec (i.e. $\sim 6.4$~kpc at $z = 0.5318$, using $H_0 = 65$~\kms~Mpc$^{-1}$ 
and an $\Omega = 1$, $\Lambda = 0$ cosmology). On the other hand, the core flux density 
was measured to be $179$~mJy at 408~MHz from a $\sim 0.6 \times 0.5$~arcsec 
resolution Merlin image (\cite{saikia90}). Using the 2.3~GHz and 408~MHz flux 
densities to estimate the core spectral index, we obtain a flux density of 
$\sim 142$~mJy at the redshifted 21cm line frequency of $\sim 927$~MHz. If the 
DLA covers only the core, the spin temperature of the absorber is $T_{\rm s} \sim 20$~K, 
a remarkably small value. However, the 21cm absorption is spread over 
$\sim 40$~km~s$^{-1}$ and is hence unlikely to arise from a single line of sight. 
Further, the 408~MHz Merlin map (\cite{saikia90}) shows that the entire flux at 
this frequency is contained within $\sim 3$~arcsecs ($\sim 16.8$~kpc at $z = 0.5318$). 
It is thus possible that the DLA covers some of the extended flux of the background 
source, besides the core. VLBI observations in the redshifted 21cm line would be 
necessary to detect the source components towards which the absorption occurs, 
and hence, the covering factor. For now, we note that a covering factor of 
unity results in $T_{\rm s} = 310~K$; this is a definite {\it upper limit} on the spin 
temperature, which must then lie between 20~K and 310~K.

9. PKS~2128$-$123 has a very flat spectrum between 8.4 GHz and 750~MHz, 
with most measurements between these frequencies yielding a flux density of
around 2~Jy (see the NED\footnote{NASA/IPAC Extragalactic Database} for details).
VLBA maps at 2.3 and 8.4 GHz show that most of this flux is contained within 
$\sim 20$~milli-arcseconds, implying that the covering factor is likely to be close to 
unity. The HI column density of the $z \sim 0.4298$ absorber is $\NHI = 2.5 
\pm 0.06 \times 10^{19}$~\cm~(\cite{ledoux2002}); the $3\sigma$~lower limit on the spin 
temperature is then $T_{\rm s} (3 \sigma) > 980$~K ($\Delta V = 7$~\kms).

10. Finally, TXS~2342+342 has a fairly flat spectrum between 1.4~GHz and 8~GHz 
(\cite{wilkinson98}; \cite{white92}), indicating that the quasar is very
compact at these frequencies. However, as Carilli et al. (1996) point 
out, the spectrum rises slowly down to the redshifted 21cm frequency of 
356~MHz, indicating the presence of a more extended source component. 
Carilli et al. (1996) argue that the DLA is likely to at least cover 
the compact component and the covering factor is hence larger than 0.5; 
it is, however, unclear whether the extended flux is also covered. We will 
use $f = 1$ in the analysis (following \cite{carilli96}); the HI column 
density of $\NHI = 2 \pm 0.5 \times 10^{21}$~\cm~ then yields $T_{\rm s} (3 \sigma) > 3585$~K 
($\Delta V = 15$~\kms). Carilli et al. (1996) estimated $T_{\rm s} > 1800$~K for 
this DLA.

\section{Discussion}
\label{sec:dis}

\begin{table*}[t!]
\label{tab:basicdata}
\begin{center}
\caption{Damped Ly-$\alpha$ Systems with 21cm Observations}
\begin{tabular}{@{}|c|c|c|c|c|c|c|c|c|c|c|}
\hline
Name & $z_{\rm abs}$ & $\NHI$ & $f$ & $ (1/f) \int \tau {\mathrm d}V$ & ${\Delta V_{21}}^{  a}$ 
& $T_{\rm s}^{  b}$ & Absorber & $b^d$& Refs.\\
  &&$10^{20}$~\cm&  & \kms & \kms &K & ID & kpc & \\
&  &&&&&&&& \\
  \hline
&  &&&&&&&&\\
  0738+313 (A)  &0.09123 &$15\pm 2$& 1   &$1.0\pm0.01$&40  &$825\pm120$& Dwarf  &$<4$ & 1,2,3 \\
PKS~0439$-$433$^c$&0.10097& ?    & 1     &$0.07\pm0.02$     &10  & ? &Spiral ?& 8.0 & 4,5   \\
  0738+313 (B)  &0.22125 &$7.9\pm1.4$& 1 &$0.488\pm0.004$&30  &$890\pm165$& Dwarf  & 20  & 1,3,6   \\
  PKS~0952+179 &0.23779 &$21\pm2.5$ &0.25 &$0.56\pm0.02$ &30&$2055\pm330$ & LSB    &$<7$ & 7,8,9  \\
PKS~1413+135$^{c\star}$&0.24671&200$^\star$& 0.12&59.43 &39&185& Spiral & ?   & 10,11,12\\
  PKS~1127$-$145&0.3127  &$51\pm9$ & 1 &$3.074\pm0.002$&120 &$910\pm160$& LSB    &$<10$& 8,9,13 \\
  PKS~1229$-$021&0.39498 &$5.6\pm0.5$& 0.5 &$1.81\pm0.01$ &110 &$170\pm15$ & Spiral &  11 & 14,15,16\\
  QSO~0248+430  &0.3941  &  ?        & ?   & 2.99     &40  & ?       &  -     & -   & 17   \\
  PKS~2128$-$123&0.4298  &$0.25\pm0.06$& 1 &$<0.014$  &-   &$>$ 980 &  -     & -   & 18,19\\
  3C196         &0.43670&$6.3\pm1.6$& ?   & ?        &250 & -       & Spiral & 9.2 & 14,20,21\\
PKS~1243$-$072  &0.4367 & ?         & 1   &0.38      &23  & ?       & Spiral & 11.3& 17,22 \\
  3C446$^c$     &0.4842 &6.3        & 0.1 &$<0.22$   &-   &$>$ 1570 & -      & -   & 13,23 \\ 
  A0~0235+164   &0.524  &50        & 1  &$13\pm0.6$  &125 &210  & Spiral &$<11.2$&  24,25,26 \\
  B2~0827+243   &0.5247 &$2.0\pm0.2$& 0.67&$0.33\pm0.03$&50  &$330\pm70$  & Spiral & 34  & 7,8,9 \\
  PKS~1629+12 &0.5318  &2.8        & 1   &$0.494\pm0.001$     &40  & 310  & Spiral & 17  & 9,19 \\
PKS~0118$-$272$^c$&0.5579  & ?         & 0.5 &$<0.12$   &-   & ?       &  -     & -   & 7,27 \\
  3C286      &0.69215 &$20\pm1.6$  &1 ?&$0.91\pm0.09$ &18 & $1205\pm240$& LSB    & 19  & 15,28,29 \\
  PKS~0454+039    &0.8596  &$5\pm0.16$ & 1   &$<0.32$   &-   &$>$855 & Dwarf  & 6.4 & 15,30,31 \\
  PKS~0215+015$^c$&1.3449  &0.8        & 1   &$<0.043$ &-   &$>$ 1020 & -      & -   &  19,32 \\
  QSO~0957+561A   &1.3911  &$2.1\pm0.5$& ?   &-        &-   & -       & -      & -   & 8,19 \\
  PKS~1354+258    &1.4205  &$32\pm2$   & -   &-        &-   & -       & -      & -   & 8,19 \\
  MC3~1331+170    &1.77636 &$15\pm1.5$ & 0.8 &$0.5\pm0.1$&25  &$1645\pm615$ & - & -   & 31,33,34 \\
  PKS~1157+014    &1.944   &$63\pm10$  & 0.55&$4 \pm 0.2$ &60 &$865 \pm190$ & - & -   & 31,35,36 \\
  PKS~0458$-$02   &2.03945 &$45\pm8$  & 1    &$6.4\pm0.4$ &30 &$385\pm100$& -      & -   & 34,37,47\\
  PKS~0528$-$2505 &2.8110  &$22\pm1$   & 1   &$<1.70$  &-   &$>$ 710  & -      & -   & 38,39,40 \\
  TXS~2342+342    &2.9084  &$20\pm0.5$ & 1   &$<0.306$ &-   &$>$ 3585 & -      & -   & 19,41 \\
  PKS~1354$-$107  &2.966   &6          & 1   &$<0.345$ &-   &$>$ 955 & -      & -   & 19,42 \\
  PKS~0537$-$286  &2.9742  &2          & 1   &$<0.058$ &-   &$>$ 1890 & -      & -   & 19,42 \\
  PKS~0336$-$014  &3.0619  &$16\pm1.3$ & 1   &$<0.095$ &-   &$>$ 9240 & -      & -   & 19,41,46\\
  PKS~0335$-$122  &3.178   &$6\pm1$    & 1   &$<0.084$ &-   &$>$ 3920 & -      & -   & 19,42\\
  PKS~0201+113    &3.3875  &$18\pm3$   & 1   &$<0.30$  &-   &$>$ 3290 & -      & -   & 41,44,45 \\
&  &&&&&&& &\\
  \hline
  \end{tabular}
\end{center}
\vskip 0.05in
  ${}^\star$~ The $z = 0.24671$ absorber towards PKS~1413+135 may be an ``associated''
absorption system, rather than an intervening one (\cite{perlman02}). 
Further, its HI column density is estimated from an X-ray spectrum (\cite{stocke92}).\\
  ${}^a$~ The velocity width $\Delta V$ quoted here is the entire velocity range
  over which absorption is seen.\\
  ${}^b$~The spin temperature values (or $3\sigma$ limits) quoted here have been
  consistently (re-)computed from the published data using Eq.~(\ref{eqn:tspin}), and the latest available values for all parameters in the
  equation. The values quoted here and the errors have been rounded off to the 
  nearest 5~K. The errors on $\NHI$ and hence, on $T_{\rm s}$, have been quoted when 
  available in the original references; in the case of asymmetric errors, the 
  larger value has been quoted. In certain instances, the values may differ 
  slightly from that quoted by the authors in the original reference. The 
  only exception is PKS~0201+113, for which multiple and quite different 
  values are quoted in the literature. We use the value from Kanekar \& 
  Chengalur (1997), modified by the new HI column density of $1.8 \pm 0.3 
  \times 10^{21}$~\cm, from Ellison et al. (2001b). \\
  ${}^c$~Candidate \dla systems, based on IUE spectra, metal lines or X-ray spectra.\\
  ${}^d$~The impact parameter $b$ has been calculated using an $\Omega = 1$, 
$\Lambda = 0$ Universe, with $H_0 = 65$~\kms~Mpc$^{-1}$.\\
\vskip 0.05in

${}^{1}$~\cite{rao98}, ${}^{2}$~\cite{chengalur99}, ${}^{3}$~\cite{cohen2001},
${}^{4}$~\cite{petitjean96}, ${}^{5}$~\cite{kanekar2001a}, ${}^{6}$~\cite{kanekar2001b},
${}^{7}$~\cite{kanekar2001c}, ${}^{8}$~\cite{rao2000}, ${}^{9}$~\cite{nestor2001},
${}^{10}$~\cite{stocke92}, ${}^{11}$~\cite{carilli92}, ${}^{12}$~\cite{perlman02}.
${}^{13}$~\cite{chengalur2000}, ${}^{14}$~\cite{kanekar2002b}, ${}^{15}$~\cite{boisse98}, 
${}^{16}$~\cite{lebrun97}, ${}^{17}$~\cite{lane2002}, ${}^{18}$~\cite{ledoux2002}, 
${}^{19}$~This paper, ${}^{20}$~\cite{briggs2001}, ${}^{21}$~\cite{cohen96}, 
${}^{22}$~\cite{kanekar2002a}, ${}^{23}$~\cite{lanzetta95}, ${}^{24}$~\cite{roberts76}, 
${}^{25}$~\cite{burbidge96}, ${}^{26}$~\cite{cohen99}, ${}^{27}$~\cite{vladilo97}, 
${}^{28}$~\cite{steidel94}, ${}^{29}$~\cite{davis78}, ${}^{30}$~\cite{steidel95}, 
${}^{31}$~\cite{briggs83}, ${}^{32}$~\cite{turnshek2002}, ${}^{33}$~\cite{wolfe79}, 
${}^{34}$~\cite{pettini94}, ${}^{35}$~\cite{briggs84}, ${}^{36}$~\cite{wolfe81}, 
${}^{37}$~\cite{wolfe85}, ${}^{38}$~\cite{carilli96}, ${}^{39}$~\cite{morton80}, 
${}^{40}$~\cite{foltz88}, ${}^{41}$~\cite{white93}, ${}^{42}$~\cite{ellison2001}, 
${}^{43}$~\cite{ellison2001b}, ${}^{44}$~\cite{kanekar97}, ${}^{45}$~\cite{briggs97},  
${}^{46}$~\cite{prochaska01}, ${}^{47}$~\cite{pw99}.

\end{table*}

	Neutral gas in the galaxy is believed to exist in two stable phases,
a cold dense phase (the cold neutral medium, CNM), with typical temperatures
of $\sim 100$~K, and a warm rarefied phase (the warm neutral medium, WNM),
with typical temperatures of $\sim 8000$~K. It should be emphasised that 
such a multi-phase medium exists only over a finite range of interstellar pressures 
(\cite{wolfire95}); at higher pressures only the CNM phase exists, while at lower 
pressures only the WNM phase is found. Further, although the multi-phase structure 
results from a balance 
between the heating and cooling rates, which themselves depend on various 
quantities such as the radiation field, the metallicity,  the dust content, 
etc., it has been shown that a two-phase pressure equilibrium is possible 
over fairly wide ranges of the above quantities (and, of course, the 
interstellar pressure; \cite{wolfire95}). On the observational front, the 
existence of neutral gas in a multi-phase medium has been established
not just for gas in the solar neighborhood, but also in high velocity
clouds (e.g. \cite{braun00}), dwarf galaxies (e.g. \cite{young97}) and
DLAs (\cite{lane2000,kanekar2001b}). In the latter case of the two DLAs at
$z=0.0912$ and $z=0.2212$ toward B0738+313, it has been explicitly
shown that the observed high spin temperatures are a consequence of 
$70 - 80$~\% of the gas being in the warm phase. In this regard, 
the multi-phase structure of at least these two systems is more like that 
of dwarf galaxies (see, for example, \cite{young97}) than large spiral disks.

	Table~3 contains a list of the 31 known candidate and 
confirmed DLAs for which 21cm observations are available in the published literature
(this is an update of the list in \cite{chengalur2000}). Note that the 
Lyman-$\alpha$ line has so far not been observed for the DLAs at $z = 0.24671$ 
towards PKS~1413+135, $z = 0.3941$ towards QSO~0248+430 and $z=0.43669$ towards 
PKS~1243$-$072; however, all three systems show strong 21cm absorption, implying 
(for reasonable values of $T_{\rm s}$) HI column densities higher than $10^{20}$~\cm~
(\cite{carilli92}; \cite{kanekar2002a}; \cite{lane2002}) and have hence been 
included in the present sample. Further, 
the $z = 0.24671$ absorber towards PKS~1413+135 may be an associated system 
(\cite{perlman02}); the HI column density listed in Table~3 is estimated from the 
deficit of soft X-rays (\cite{stocke92}). No similar direct estimate exists for the HI 
column densities of the systems towards PKS~1243$-$072 and QSO~0248+430. We also note 
that two candidate DLAs (on the basis of metal lines), at $z = 0.5579$ towards 
PKS~0118$-$272 (\cite{kanekar2001c}) and $z = 0.101$ towards PKS~0439$-$433 
(\cite{kanekar2001a}) have, for completeness, been included in this table; however, 
they will not be used in the later analysis, as their HI column densities (and hence, 
spin temperatures) are unknown. Cols.~3, 4, 5, 6 and 7 of the table contain the HI 
column density (in units of $10^{20}$~\cm), the adopted covering factor $f$, the ``true'' 
equivalent width ($(1/f) \int \tau {\mathrm d}V$), the velocity spread of 21cm 
absorption (between nulls, in \kms) and the spin temperature. Besides these, Cols.~8 
and 9 list the nature of the objects identified as the DLA hosts and their impact 
parameters ($b$, in kpc) to the QSO line of sight. Of course, it is possible that 
the latter identifications are in error in some cases if, for example, the absorption 
arises in an object closer to the QSO but too faint to be identified in optical/IR 
images. With this caveat in mind, we note that seven of the DLA hosts have been 
identified with large, luminous galaxies, with luminosities close to $L_\star$.

\begin{figure}
\centering
\epsfig{file=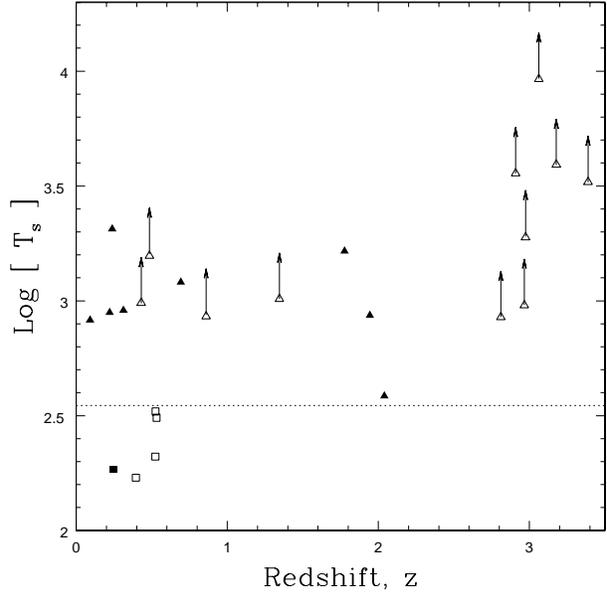,height=3.3truein,width=3.3truein}
\caption{The spin temperature (Log[$T_{\rm s}$]) of the 24 DLAs of the complete sample 
with $T_{\rm s}$ estimates as a function of redshift $z$. Note that seven DLAs listed 
in Table~3 have not been included here for reasons listed in the text. Objects 
identified with spiral galaxies are plotted using squares; the filled square 
shows the $z = 0.24671$ spiral towards PKS~1413+135, which may be an associated 
system. Other detections are shown as filled triangles, with non-detections shown 
as open triangles, with arrows. }
\label{fig:tvsz}
\end{figure}

\begin{figure}
\centering
\epsfig{file=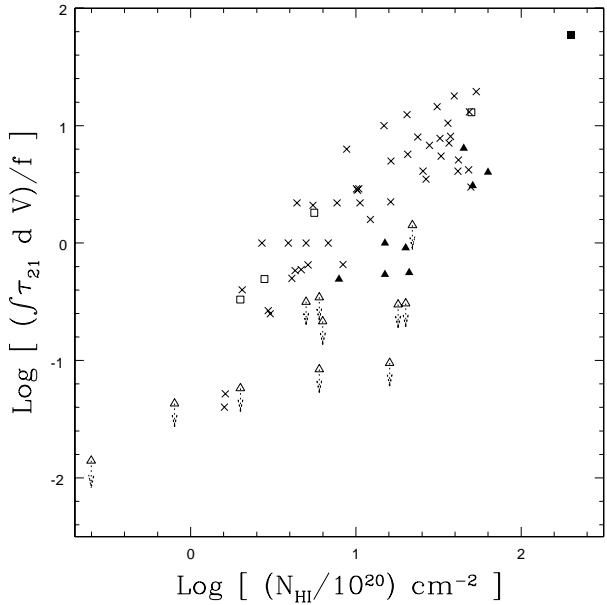,height=3.3truein,width=3.3truein}
\caption{The {\it true} equivalent width ($\int \tau \mathrm{d} V$/$f$) as a 
function of the HI column density for the 24 DLAs of the complete sample for 
which we can calculate $T_{\rm s}$ values. Objects identified with spiral galaxies are
plotted using squares; the filled square shows the $z = 0.24671$ spiral towards 
PKS~1413+135, which may be an associated system. Other detections are shown as 
filled triangles, with non-detections shown as open triangles, with arrows. The crosses 
show measurements in the Galaxy by Colgan et al. (1988) and Payne et al. (1982), using 21cm 
emission/absorption studies.}
\label{fig:eq_wid}
\end{figure}

\begin{figure}
\centering
\epsfig{file=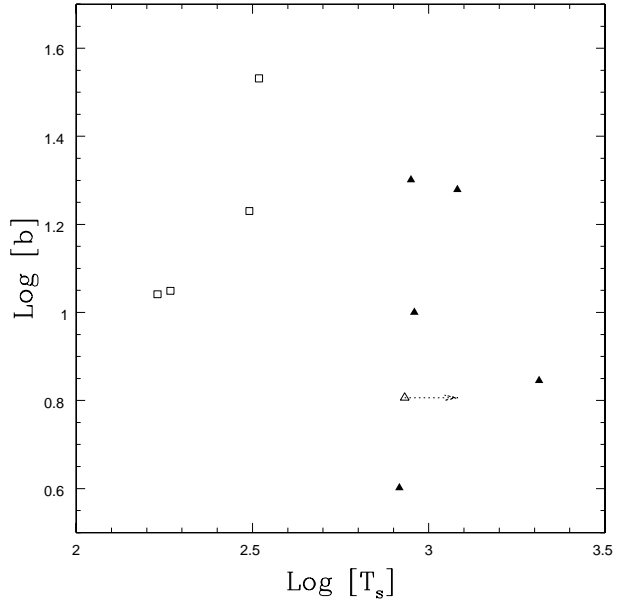,height=3.3truein,width=3.3truein}
\caption{The impact parameter $b$ (in kpc) as a function of spin temperature for 
10 DLAs at $z < 1$. The open squares represent objects identified with 
spiral galaxies; the remaining detections are shown as filled triangles, with 
non-detections shown as open triangles, with arrows. }
\label{fig:ts-b}
\end{figure}

Fig.~\ref{fig:tvsz} shows a plot of the measured spin temperatures as 
a function of redshift for the 24 DLAs of the complete sample which have reliable 
estimates of the spin temperature (note that this plot includes the $z = 0.24671$ absorber 
towards PKS~1413+135 which may be an associated system). The seven systems 
in Table~3 which are not included in Fig.~\ref{fig:tvsz} are the two 21cm absorbers 
towards PKS~1243$-$072 and QSO~0248+430 for which the 
HI column density is unknown (see above), the DLAs towards PKS~1354+258 and 
QSO~0957+561 for which the covering factor is highly uncertain (see 
Sect.~\ref{sec:Ts}), the two candidate DLAs towards PKS~0118$-$272 and PKS~0439$-$433 
and, finally, the $z = 0.437$~absorber towards 3C196, where the lines of sight to the 
optical QSO and the radio continuum trace very different paths through the absorbing 
galaxy. The dotted line in Fig.~\ref{fig:tvsz} is at 350~K -- more than 80~\% 
of spin temperature measurements in M~31 and the Milky Way have values $< 350$~K 
(\cite{braun92}). It can be seen that the spin temperatures of the majority of 
DLAs are considerably higher than those typical of lines of sight through the 
disks of the Milky Way or nearby spiral galaxies. The only exceptions to the above are 
the five systems at low $z$ which have been identified with spiral disks and 
the DLA at $z \sim 2.04$ towards PKS~0458$-$020, for which $T_{\rm s} \sim 385\pm 100$~K; we 
discuss the latter separately below. Note that it is not presently possible to estimate 
the spin temperatures of the remaining two $L \sim L_\star$ DLAs at $z \sim 0.437$ 
towards PKS~1243$-$072 and 3C196. Finally, the 21cm absorption profile for PKS~1629+12 
reported in this paper continues the  trend (noted earlier in Chengalur \& Kanekar (2000) 
and Kanekar \& Chengalur (2001)) of low spin temperatures being found in all DLAs identified 
as $L \sim L_\star$ galaxies. 

Fig.~\ref{fig:eq_wid} presents the data in an alternative way, as a 
scatter plot of the ``true'' HI 21cm equivalent width (i.e. $(1/f) \int \tau 
{\mathrm d} V$) against the HI column density.  Both of these are directly
observable quantities. The crosses show the same quantities for Galactic HI
clouds (\cite{colgan88}; \cite{payne82}), measured using 21cm 
absorption/emission studies. HI column density estimates using observations 
of the Lyman-$\alpha$ line have been found to be in good agreement with 
estimates from 21cm emission studies (\cite{dickey90}). Fig.~\ref{fig:eq_wid}
hence allows a direct comparison of the same observables for
DLAs and galactic clouds. It can be clearly seen that, at a given HI column 
density, the majority of DLAs tend to have smaller 21cm equivalent widths than those
found for lines of sight through the Galaxy. The only exceptions are DLAs 
that are known to be associated with spiral galaxies (shown as squares) 
and the $z \sim 2.04$ DLA towards PKS~0458$-$020,  (which we discuss in 
more detail below), which do lie within the range of Galactic values. The 
conclusion would seem to be that the majority of DLAs contain larger fractions 
of the warm (and thus weakly absorbing) phase of neutral hydrogen than typical 
in nearby spirals, as earlier pointed out by Carilli et al. (1996) and Kanekar 
\& Chengalur (2001).

It is possible that the high $T_{\rm s}$ values seen in DLAs might arise due to 
absorption originating in the outskirts of large disks, which would also 
provide a higher absorption cross-section than the inner regions of these 
galaxies. Fig.~\ref{fig:ts-b} shows a plot of $T_{\rm s}$ versus impact parameter 
$b$ for the 10 DLAs of the sample (all with $z < 1$) with optical identifications,
as well as estimates of both $T_{\rm s}$ and impact parameter; absorbers identified with 
spiral galaxies are shown as open squares. No correlation can be seen 
to exist between $T_{\rm s}$ and $b$. On the contrary, the largest impact parameter 
of the sample ($b = 34$~kpc) is seen in a low $T_{\rm s}$ system, the $z = 0.5247$ DLA 
towards B2~0827+243, while four of the high $T_{\rm s}$ DLAs have $b < 10$~kpc. The low 
redshift sample clearly does not support the hypothesis that the high $T_{\rm s}$ seen in
low redshift DLAs is because the absorption arises from lines of sight which pass
through the outer regions of disk galaxies. Instead, it appears that 
the spin temperature correlates with the nature of the absorbing galaxy. We discuss 
next the $z \sim 2.04$ DLA towards PKS~0458$-$020, the only high $z$ DLA with a 
spin temperature $T_{\rm s} < 400$~K.

The $z = 2.03945$ DLA towards PKS~0458$-$020 is as yet the highest redshift DLA 
with a confirmed detection of 21cm absorption (\cite{wolfe85}). Briggs et al. (1989) 
found the VLBI 21cm absorption profile to be very similar to the single dish profile 
and used this to argue that the absorber shows little structure on the scale of
the extended radio continuum emission of the background source; these observations 
hence imply that the transverse size of the absorber is larger than $8\; h^{-1}$~kpc.
Further, the low ionization metal lines of the DLA are clearly asymmetric (\cite{pw97}), 
which can be explained if the absorber is a rapidly rotating disk (although such 
asymmetries are also explanable in other models; see, for example, \cite{haehnelt98}, 
\cite{mcdonald99}). The spin temperature of the absorber was originally estimated to 
be $T_{\rm s} = 594 \pm 297 $~K (\cite{wolfe85}), the large error arising due to uncertainties 
in fitting the Lyman-$\alpha$ profile ($\NHI = 8 \pm 4 \times 10^{21}$~\cm; \cite{wolfe85}). 
Recently, however, the HI column density has been measured far more accurately by 
Pettini et al. (1994), who quote $\NHI = 4.5 \pm 0.8 \times 10^{21}$~\cm. Combining this 
with the HI 21cm absorption profile (and taking $f = 1$, as confirmed by the 
VLBI observations) yields $T_{\rm s} = 385 \pm 100$~K, 
a value not too far from those obtained in local spirals; this is the first case of a 
DLA at $z \ga 0.6$ with a spin temperature less than $400$~K. Interestingly enough, 
the 21cm equivalent width of the absorber too lies close to the range of Galactic 
values shown in Fig.~\ref{fig:eq_wid}. We note, however, that the metallicity of the 
absorber is far lower than solar values ([Zn/H] = $-1.16\pm 0.09$; \cite{pw99}), 
implying that the DLA has not undergone much chemical evolution. Detailed analysis 
of the formation of a multi-phase medium in proto-galaxies (\cite{spaans97a}; 
\cite{spaans97b}) indicate that the epoch of formation of a multi-phase ISM is a strong 
function of the galactic mass; galaxies with baryonic mass $\sim 5 \times 10^{11}$~M$_\odot$ 
form the cold phase by $ z \sim 3$ while dwarf galaxies, with baryonic mass  
$\sim 3\times 10^9$~M$_\odot$, typically form the cold phase only by $z\sim 1$. The
relatively low measured spin temperature in the $z \sim 2.04$ DLA thus makes it likely that 
the absorber is a large galaxy.  The VLBI observations and the low spin temperature
are hence both consistent with the DLA being a massive galaxy, with the high mass 
providing a sufficiently high central pressure to compensate for the low
metallicity and produce a sizeable fraction of the cold phase of HI.

\begin{table}
\label{table:cnm}
\begin{center}
\caption{The CNM fraction in high $z$ DLAs.}
\vskip 0.1in
\begin{tabular}{@{}|l|c|c|c|c|}
\hline
Name & $z_{\rm abs}$ & $\NHI $ & $\NHI $ & $f_{\rm CNM}$ \\
&& (200~K)&(Total) & \\
&&$ \times 10^{20}$& $\times  10^{20}$& \\
&&&& \\
\hline
&&&& \\
PKS~0528$-$250  & 2.8110 & $<$6.2  & $22 \pm 1$  & $<$0.28 \\
TXS~2342+342    & 2.9084 & $<$0.7  & $20\pm0.5$  & $<$0.03 \\
PKS~1354$-$107  & 2.966  & $<$1.3  & 6.0         & $<$0.21 \\
PKS~0537$-$286  & 2.9742 & $<$0.2  & 2.0         & $<$0.09 \\
PKS~0336$-$014  & 3.0619 & $<$0.2  & $16 \pm1.3$ & $<$0.01 \\
PKS~0335$-$122  & 3.178  & $<$0.2  & $6\pm1 $    & $<$0.04 \\
PKS~0201+113    & 3.3875 & $<$1.1  & $18 \pm 3$  & $<$0.06 \\
&&&& \\
\hline
\end{tabular}
\end{center}
\end{table}

	The evolution of spin temperature with redshift has been the subject 
of interest for some time. The original observations of high spin temperatures 
at high redshift led to the suggestion that the elevation of the spin 
temperature was perhaps due to evolutionary effects. More recently, 
the detection of low redshift DLAs ($z \la 0.2$) with high $T_{\rm s}$ values 
(\cite{lane98,chengalur99}) indicated that a simple trend did not exist in 
the evolution of $T_{\rm s}$ with redshift. However, Chengalur \& Kanekar (2000) noted the 
following trend: DLAs with both low and high spin temperatures are found at
low redshifts ($z \la 1$), while only high $T_{\rm s}$ DLAs are found at high redshifts 
($z \ga 3$). The present, larger sample of Fig.~\ref{fig:tvsz} shows this trend 
clearly, with the exceedingly high values of spin temperature at $z \ga 3$ indicating 
that these systems contain very small fractions of the CNM phase of HI. 

The non-detection of HI absorption in the present fairly deep search can be used 
to place stringent constraints on the fractional content of CNM in the high $z$ 
absorbers. To do so, we note that the allowed values of CNM temperatures in the 
Milky Way lie between 40~K and 200~K (\cite{wolfire95}). We hence use 200~K (in 
Eq.~(\ref{eqn:tspin})) as an upper limit to the temperature of the CNM phase 
in the seven $z \ga 3$ DLAs to estimate upper limits on the CNM fraction in these 
systems. The original high resolution spectra (see Table~1) are used for this 
purpose, not the smoothed spectra used to estimate the spin temperature. 
The results are shown in Table~4 where the third column gives the $3\sigma$ upper 
limit to the HI column density in 200~K gas (note that this limit 
scales linearly with the assumed CNM temperature and would hence be even lower for 
lower temperature CNM). The last column in this table gives the upper limit to the 
fraction of the HI in CNM, $f_{\rm CNM}$; it can be seen that five of the seven systems 
at $z \ga 3$ have CNM fractions less than 10~\%, with three of these having 
$f_{\rm CNM} < 5$~\%. Clearly, most of the gas in the high $z$ DLAs is warm, 
the typical fraction of CNM gas is even smaller than that in the two high 
$T_{\rm s}$ DLAs at low redshift with measured CNM fractions (\cite{lane2000,kanekar2001b}) 
or in local dwarf galaxies (\cite{young97}).

An independent line of evidence for high temperatures at high redshift comes from 
observations of the $n$(CII)/$n$(CI) ratio (\cite{liszt2002}), which is quite different 
in the CNM and the WNM. Liszt (2002) used observations of this ratio in 11 DLAs to 
suggest that high redshift ($z>2.3$) DLAs are dominated by warm gas, with at most 
a few per cent of the neutral HI in the CNM. This is in agreement with our results. 
Besides this, it has been observed that high $z$ DLAs typically have very low 
molecular gas fractions; this can be understood in models in which the absorbing 
gas is at high temperature (\cite{petitjean2000}; but see also \cite{liszt2002}). 
We note, in passing, that it is exceedingly interesting that the $z = 3.0619$ DLA 
towards PKS~0336$-$014 has $T_{\rm s} > 9240$~K, implying that its gas is significantly hotter 
than even the standard WNM in the Milky Way (which has temperatures in the range 
$5000 - 8000$~K; \cite{wolfire95}). 

	Of the 24 DLAs with $T_{\rm s}$ measurements, 16 are at $z < 2$ and 8 at $z > 2$; we
will call these the low and high $z$ samples respectively. The sample is of a sufficiently
large size to carry out a statistical comparison to test the 
likelihood that the systems in the two samples stem from the same parent distribution. 
The Gehan test (\cite{gehan65}; \cite{miller81}) is appropriate for this comparison 
as it can be directly used in the case of limits on the measured quantity (in our 
case, the spin temperature). When applied to our data, this test rules out the possibility 
of the two samples being drawn from the same 
distribution at the $\sim 99$\% confidence level. (If one chooses to treat the $3\sigma$
upper limits as detections, a Kolmogorov-Smirnov rank 1 test rules out the above 
hypothesis at the 95~\% confidence level.) It is thus very unlikely that the high and 
low $z$ DLAs are systems of similar ``character''. We note that $z = 2$ was used 
as the redshift of separation of the two samples as this is the approximate redshift 
where star formation is believed to peak; it is hence plausible that systems at $z > 2$ 
will show differences from those at lower redshifts. Since we have only 4 systems with 
$T_{\rm s}$ estimates in the redshift range $1 < z < 2.8$ (the $z \sim 1.4$ DLAs 
towards PKS~1354+258 and QSO~0957+561A do not have good estimates of $T_{\rm s}$), it is 
not possible to independently determine the threshold redshift with the present sample.

In conclusion, the sample of damped Lyman-$\alpha$ systems which have been searched 
for 21cm absorption now consists of 31 systems; estimates of the spin temperature 
are available in 24 cases, of which 16 DLAs are at $z < 2$ and 8 at $z > 2$. All 
of the low $T_{\rm s}$, low $z$ DLAs have been identified with large, luminous galaxies, while all 
DLAs at $z < 1$ with high spin temperature ($T_{\rm s} \ga 1000$~K) have been identified 
either with LSBs or dwarfs. It thus appears likely that the spin temperature depends
mainly on the nature of the absorbing galaxy, as suggested earlier by Chengalur \& 
Kanekar (2000). We use a recent measurement of the HI column density in the $z = 2.03945$ DLA 
towards PKS~0458$-$020 to derive a relatively low $T_{\rm s}$ value in the absorber; this is 
the first high $z$ DLA to show $T_{\rm s} < 400$~K. The low metallicity allied with a 
low spin temperature suggest that the DLA is a massive galaxy, consistent
with VLBI limits of its transverse size. The new deep 
GMRT 21cm spectra of this paper have also been used to estimate the fraction of HI 
in the cold phase, $f_{\rm CNM}$, in the absorbers at $z \ga 3$; it was found that 
$f_{\rm CNM} < 0.3$ in all seven absorbers, with $f_{\rm CNM} < 0.1$ in five of the seven 
cases. These results are in good agreement with estimates of $f_{\rm CNM} \sim few$~\%, 
from observations of the $n$(CII)/$n$(CI) ratio (\cite{liszt2002}). Finally, we have 
used the above spin temperature estimates to test the likelihood that the $z > 2$ 
and $z < 2$ DLAs are drawn from the same parent population and rule out this 
hypothesis at the $\sim 99$\% confidence level.

\begin{acknowledgements}
We thank the referee, Wendy Lane, for numerous suggestions and comments on an 
earlier version which have significantly improved this paper. These observations 
would not have been possible without the many years of dedicated effort put in 
by the GMRT staff in order to build the telescope. The GMRT is operated by the 
National Centre for Radio Astrophysics of the Tata Institute of Fundamental Research.
This research has made use of the United States Naval Observatory (USNO) Radio 
Reference Frame Image Database (RRFID) and the NASA/IPAC Extragalactic Database 
(NED) which is operated by the Jet Propulsion  Laboratory, California Institute 
of Technology, under Contract with the National Aeronautics and Space Administration.

\end{acknowledgements}

\end{document}